\documentclass[authoryear, 12pt]{elsarticle}
\usepackage[english]{babel}

\usepackage[toc,page]{appendix}
\usepackage{exscale}
\usepackage{amssymb}        
\usepackage{amsthm,amsmath}        
\usepackage{natbib}         
\usepackage{mathtools}
\usepackage{graphicx}       
\usepackage{subfigure}  
\usepackage{tabularx, booktabs, multirow}    
\usepackage{color}
\usepackage{gensymb} 
\usepackage{float} 
\journal{Materials Science and Engineering A, accepted for publication.}

\title{Atomistic simulations of the interaction of basal dislocations with MgZn$_2$ precipitates in Mg alloys}

\author{G. Esteban-Manzanares$^{1, 2}$}
\author{R. Alizadeh$^{1}$}
\author{I. Papadimitriou$^{1}$}
\author{D. Dickel$^{3}$}
\author{C. D. Barrett$^{3}$}
\author{J. LLorca$^{1, 2, }$\corref{cor1}}
\address{$^1$ IMDEA Materials Institute, C/ Eric Kandel 2, 28906, Getafe, Madrid, Spain. \\  $^2$ Department of Materials Science, Polytechnic University of Madrid/Universidad Polit\'ecnica de Madrid, E. T. S. de Ingenieros de Caminos. 28040 - Madrid, Spain. \\  $^3$ Department of Mechanical Engineering, Mississippi State University, MS 39762, USA.}

\cortext[cor1]{Corresponding author}

\begin{document} 

\begin{abstract}

The interaction between Mg edge basal dislocations and rod-shaped $\beta_1'$-MgZn$_2$ precipitates was studied  by atomistic simulations using a new interatomic potential. The atomistic model was carefully built taking into account the experimental information about the orientation relationship between the matrix and the precipitate to ensure minimum energy interfaces. It was found that the dislocations initially overcame the precipitate by the formation of an Orowan loop that penetrated in the precipitate. The precipitate was finally sheared after several Orowan loops were piled-up. The number of loops necessary to shear the precipitate decreased as precipitate cross-section decreased and the temperature increased but was independent of the precipitate spacing. Precipitate shearing did not take place along well-defined crystallographic planes but it was triggered by the accumulation of the elastic energy in the precipitate which finally led to formation of an amorphous layer below and above the slip plane of the basal dislocations. The kink induced in the precipitate by this mechanism was in good agreement with transmission electron microscopy observations.

\end{abstract}

\begin{keyword}
Mg-Zn alloys, precipitation strengthening, atomistic simulations, precipitate shearing.
\end{keyword}

\maketitle

\section{Introduction} \label{Intro}

Dispersion of nm-sized intermetallic precipitates is one of the most efficient strategies to increase the strength of metallic alloys \citep{N97}. Precipitate hardening depends on the size, shape, volume fraction and spatial distribution of the precipitates in the matrix as well as on the mechanisms of dislocation-precipitate interaction. Large precipitates, normally incoherent with respect to the matrix, are usually considered impenetrable for dislocations that have to overcome the obstacle by the formation of an Orowan loop around the precipitate \citep{O48}. The Orowan model captures very well the physics of dislocation/precipitate interactions in the case of impenetrable precipitates and has been refined over the years (within the framework of continuum models) to take into account the attraction between opposite dislocation segments during the formation of the loop \citep{BKS73} as well as of the shape, size and spatial distribution of the precipitates \citep{A85, N97, N03, N12}. More recently, dislocation dynamics simulations have been used to account for the influence of the elastic mismatch, precipitate shape and orientation, the presence of eigenstrains, etc. on the CRSS to overcome the precipitate \citep{XSC04, QMD10, MND11, GFM15, ZSD17, SEP18}. 

On the contrary, small precipitates coherent with the matrix tend to be sheared by dislocations. This is the case of Guinier-Preston (GP) zones in many metallic alloys \citep{BFK61, RBP18}, $\theta''$ and $\Omega$ precipitates in Al-Cu and Al-Cu-Mg alloys \citep{N14}, and $\gamma'$-Ni$_{3}$(Al, Ti)  and $\gamma''$-Ni$_3$Nb in Ni-based superalloys \citep{XCC08, CGJ15}, etc. In general, the critical resolved shear stress (CRSS) required to shear a precipitate is a function of different factors, namely chemical energy, stacking fault energy, modulus mismatch, coherency strains and order strengthening \citep{A85, N97}. All these strengthening mechanisms can appear simultaneously in engineering alloys and the different contribution of each one to the CRSS has been evaluated using continuum models based on the dislocation line tension \citep{A85, N97} or, more recently, through dislocation dynamics simulations \citep{VDR09, GFM15, HZT12, HRU17}. These strategies were  able to capture the experimental trends in the case of Ni-based superalloys, although direct comparisons were more difficult because some of the critical parameters for these models (stacking fault energy of the precipitate, coherency strains, antiphase boundary energy) are not always easy to estimate. Moreover, the validity of these approaches can be questioned when the continuum hypothesis is no longer applicable due to the small dimensions of the precipitate and atomistics simulations are becoming more useful to understand these processes \citep{EMS19, EBM19}.

In general, precipitate shearing is not as efficient as the formation of Orowan loops to strengthen metallic alloys because successive shearing of the precipitates tends to reduce the  size of the precipitates up to a point in which they can no longer offer any resistance to the movement of dislocations. As a result, deformation is localized in soft slip bands free of precipitates \citep{XCC05}. On the contrary, significant strain hardening is promoted in the presence of Orowan loops when successive dislocations try to overcome the precipitate because the strong repulsion with the Orowan loop around the precipitate leads to an increase in the CRSS. 

Recent experimental investigations have reported a mixture of Orowan loops and precipitate shearing in several Mg alloys containing large precipitates \citep{ZWZ18, HYQ19, CCP19, AL19} and the competition between both mechanisms has been shown by the atomistic simulations of \cite{EMP19} in the case of Mg alloys strengthened with Mg$_{12}$Al$_{17}$ precipitates. It was argued that the Mg basal plane is always parallel to one crystallographic plane of the precipitate ($(0001)_{Mg} \| (110)_{\beta}$ in Mg-Al;  $(0001)_{Mg} \| (110)_{\beta_1}$ in Mg-Nd) and that, even though the matrix/precipitate interfaces may be incoherent, basal dislocations in the Mg matrix can glide into the precipitate without changing the slip plane. Of course, the CRSS to shear the precipitate is much higher than that for basal slip in Mg and several basal dislocations have to pile-up at the interface before the precipitate is sheared. Thus, precipitate shearing by basal dislocations hinders precipitation hardening in Mg alloys.

In this investigation, starting from the experimental evidence provided by transmission electron microscopy observations, a realistic atomistic model of the rod-shaped MgZn$_2$ precipitates embedded in the Mg matrix was created using a new interatomic potential which was validated for this system. Afterwards, the mechanisms of dislocation/precipitate interaction were analysed by means of molecular statics and molecular dynamics simulations and compared with those observed experimentally. The atomistic simulation results showed light about the influence of precipitate dimensions, spacing and temperature on the CRSS.

\section{Experimental evidence}\label{EE}

\begin{figure}[!]
	\centering
\includegraphics[width=\textwidth]{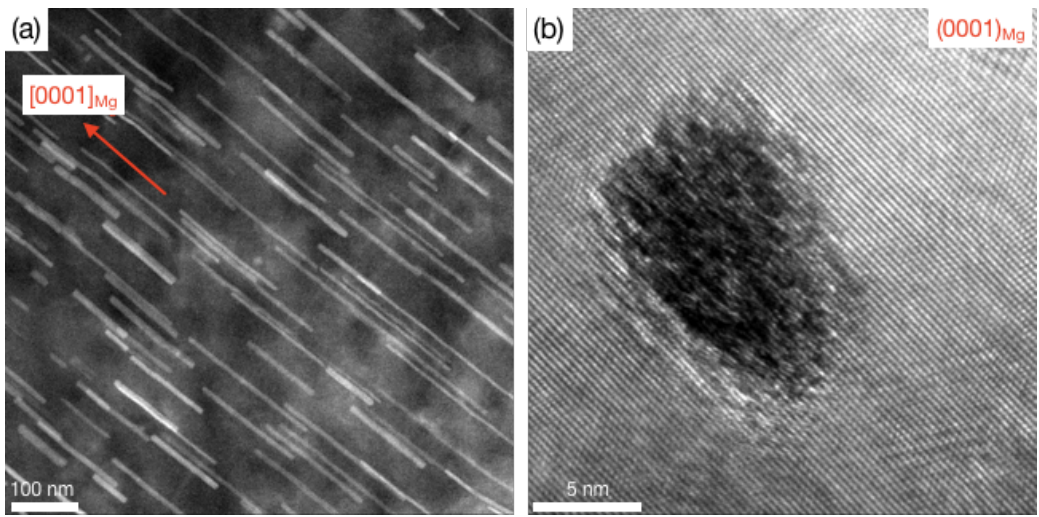}
	\caption{Scanning transmission electron micrographs of a lamella of the Mg - 4 wt.\% Zn alloy aged at 149 $^\circ$C for 100 hours. (a)   Dark-field micrograph with the lamella perpendicular to the prismatic plane of Mg. (b) High resolution micrograph with the lamella parallel to the basal plane of Mg.}
	\label{precipitates}
\end{figure}

A Mg-4 wt.\% Zn alloy was manufactured by casting from high purity Mg (99.90 wt. \%) and Zn (99.99 wt. \%) pellets. The cast rods, were homogenized at 450 $^\circ$C for 15 days in Ar atmosphere. Afterwards,  they subjected to a solution heat treatment at 450 $^\circ$C for 20 days in Ar atmosphere and aged at 149$^\circ$C for 100 hours.  The microstructure of the aged alloy was analysed by transmission electron microscopy and showed a homogeneous distribution of elongated precipitates parallel to the $c$ axis of the Mg matrix, Fig. \ref{precipitates}$(a)$. The shape of the precipitates in the basal plane of Mg was more or less equiaxed, Fig. \ref{precipitates}$(b)$. The average length and diameter of the precipitates were 146 $\pm$ 10 nm and 9.7 $\pm$ 2 nm as measured from scanning transmission electron micrographs of lamellae perpendicular to the prismatic and basal planes of the Mg matrix, respectively,  using ImageJ. They were similar to those reported by \cite{WS15} in a Mg-5 wt. \% Zn subjected to similar aging conditions and have been identified as $\beta'_1$ precipitates with the structure of the hexagonal MgZn$_2$ Laves phase \citep{C65, CB69}. The lattice parameters of the $\beta'_1$ phase are  $c$ = 0.85 nm and $a$ = 0.52 nm \citep{CB69}  and the latter is very close to the $c$ lattice parameter of the Mg matrix (0.52 nm). Thus, the  $\beta'_1$ precipitates grow along $[2\bar 1 \bar 1 0]_{\beta'_1} \parallel [0001]_{Mg}$ while the compact basal plane of Mg is parallel the prismatic plane of the precipitate, i.e.  $(2\bar 1 \bar 1 0)_{\beta'_1}\parallel (0001)_{Mg}$ \citep{SRS10}. Nevertheless, the interfaces between the Mg and the precipitate parallel to the $c$ axis of Mg are not known and  the irregular shape of the precipitate cross-section seems to indicate that several different interfaces are formed although the high resolution transmission electron micrograph in Fig. \ref{precipitates}$(b)$ is compatible with the hypothesis that one of them is $(2\bar 1 \bar 1 0)_{Mg}\parallel (0001)_{\beta'_1}$.

\begin{figure}[!]
	\centering
	\includegraphics[width=\textwidth]{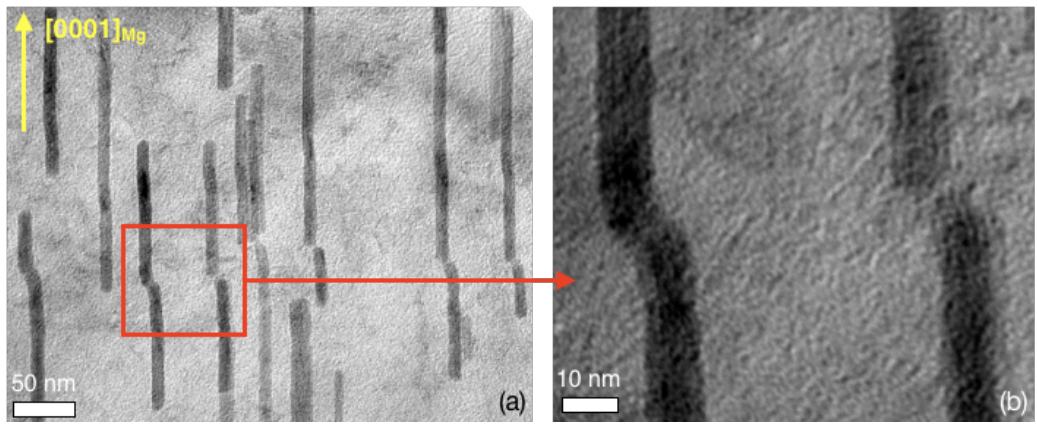}
	\caption{Bright-field transmission electron micrographs showing the shearing of the precipitates by basal dislocations in a Mg-4 wt.\% Zn alloy aged at 149 $^\circ$C for 100 hours. (a) Large view showing various precipitates sheared by basal dislocation. (b) Detail of precipitate shearing showing the formation of a kink in the precipitate before shearing.}
	\label{DPinteraction23}
\end{figure}

Micropillars of  5 x 5 $\mu$m$^2$ square cross-section were manufactured by focused ion beam from grains suitable oriented for basal slip and tested in compression using a nanoindenter with a flat punch at 23 $^\circ$C. Thin lamellae were extracted from the deformed micropillars by focused ion beam and analysed in the transmission electron microscope to ascertain the deformation mechanisms. They are shown in Fig. \ref{DPinteraction23} for the micropillar deformed at 23$^\circ$C, which presents a clear evidence of precipitate shearing by basal dislocations, in agreement with previous experimental observations \citep{WS15}. Nevertheless, the high magnification micrographs in Fig. \ref{DPinteraction23} clearly show that shearing of the precipitate did not take place along one single crystallographic plane but was associated with the formation of a kink along the precipitate. This mechanism indicates that the precipitate was subjected to large elastic stresses prior to shearing, which were very likely associated to the formation of dislocation pile-ups at the matrix/precipitate interface. In fact, detailed analysis of the deformation mechanisms by transmission electron microscopy in this alloy also showed the presence of dislocations bowing between the $\beta'_1$ precipitates \citep{AL19}.

\section{Interatomic potential validation}\label{IP}

A second nearest-neighbour modified embedded atom method (MEAM) potential was used to evaluate dislocation-$\beta_1'$ precipitate interactions. It was based in the Mg-Al-Zn ternary MEAM potential developed by \cite{DD18}, which provides good predictions for the stacking fault energy and Peierls stress of basal dislocations in Mg and was recently used to simulate the interaction of basal dislocations in Mg with $\beta$-Mg$_{12}$Al$_{17}$ precipitates \citep{EMP19}. The Mg-Zn interaction parameters in this potential were modified to improve the predictions of the elastic constants of the $\beta_1'$ precipitates as well as the interface energies of the main interfaces between the Mg and $\beta_1'$ lattices with those obtained using density functional theory (DFT). 
The new parameters of the modified MEAM can be found in the appendix \ref{AppMP}.

DFT calculations were carried out using Quantum Espresso \citep{qe}, while LAMMPS \citep{lammps} was employed in all the molecular mechanics calculations. The same simulation strategy was employed to determine the elastic constants MgZn$_2$-$\beta'_1$ precipitate using molecular statics (MS) and DFT.
 
A given strain was prescribed to the atomistic domain and the internal coordinates of the atoms were relaxed. The stresses were determined once the stable state at given strain was attained. The unit cell was deformed in different directions taking into account the symmetries of the hexagonal crystal to obtain all the independent elastic constants. Six strain steps, from -0.003 to 0.003, were used for each direction to adjust a linear stress-strain relationship. The non-linear conjugate gradient algorithm (CG) with force tolerance  1$\times$10$^{-4}$ eV/$\textrm{\AA}$ was used in the MS simulations. DFT calculations were carried out using the Perdew-Burke-Erzenhof approach \citep{GGA} to examine the exchange-correlation energy, within the Generalized Gradient Approximation. The basis set of plane wavefunctions, used to describe the real electronic functions, were reduced by means of ultrasoft pseudopotentials \citep{USP}. A cut-off of 36 Ry (490 eV) was found to be accurate enough, assuming an error in the total energy of less than 1 meV/atom. A separation of 0.03 $\textrm\AA^{-1}$ in the k-point grid was employed and the integration over the Brillouin zone was carried out according to the Monkhorst-Pack scheme \citep{MP}.  

 The elastic constants of $\beta'_1$ precipitate obtained via both atomistic procedures are shown in Table \ref{Beta1_EC}, together with experimental data available in the literature. It is readily observed that the modifications of the MEAM potential in \cite{DD18} lead to a much closer agreement of the elastic constants predicted by MS with both DFT and experimental data \citep{Seidenkranz1976}.

\begin{table}[t]
 \begin{center}
    \caption{Comparison of the elastic constants (GPa) of $\beta'_1$-MgZn$_2$ precipitate obtained by atomistic simulations and experiments.} 
    \label{Beta1_EC}
    \begin{tabular}{l|c|c|c|c|c} \hline \hline
      Method & $C_{11}$ & $C_{33}$ & $C_{12}$ & $C_{13}$ &  $C_{44}$ \\
      \hline
		MS (this work) & 107.2 & 124.2 & 37.2 & 33.4 & 20.6 \\
		DFT (this work) & 107.7 & 124 & 49.4 & 33.3 & 25.1 \\
		MS \citep{DD18} & 145.8 & 164.5 & 48.9 & 43.6 & 36.3 \\
		Exp. \citep{Seidenkranz1976} & 107.3 & 126.4 & 45.5 & 27.4 & 27.7 \\
      \hline \hline
    \end{tabular}
  \end{center}
\end{table}

\begin{figure}[!t]
	\centering
	\includegraphics[width=\textwidth]{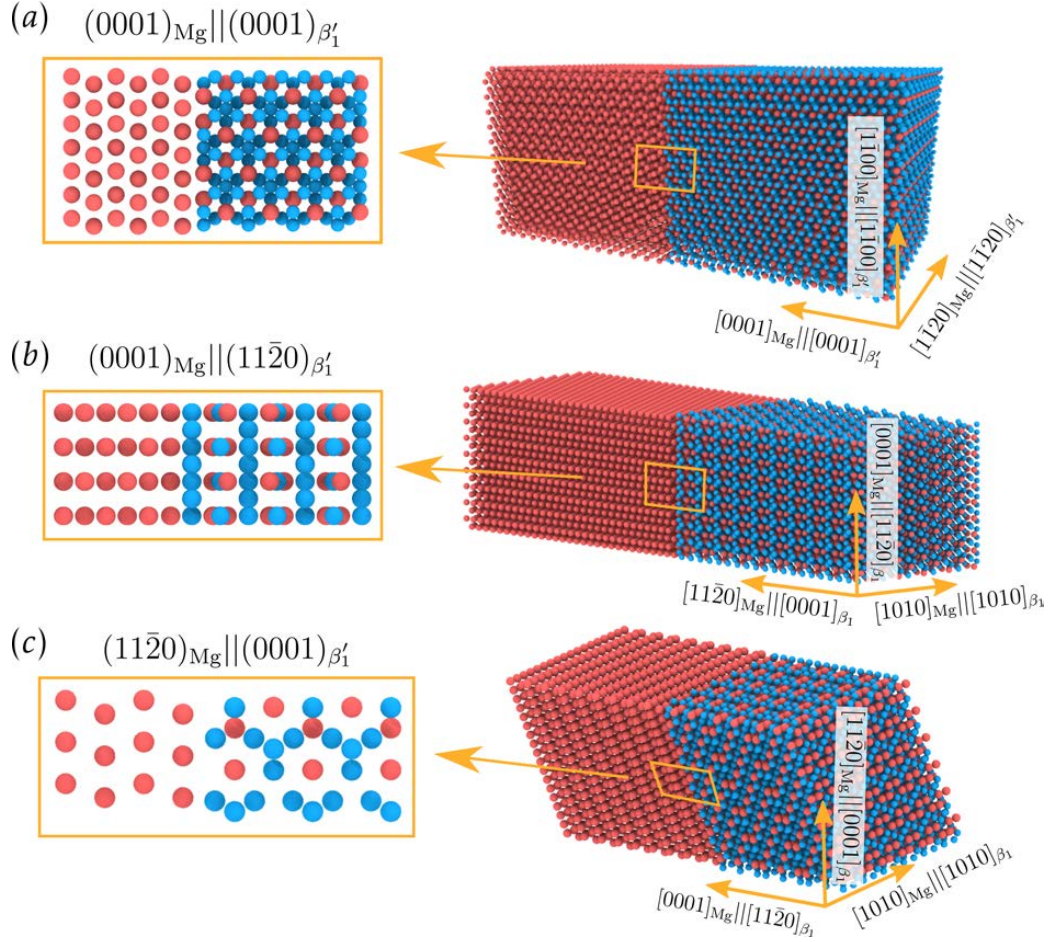}
	\caption{Atomistic supercells used for the calculation of the interfaces energies. $(a)$ $(0001)_{\text{Mg}}\parallel(0001)_{\beta'_1}$. $(b)$ $(0001)_{\text{Mg}}\parallel(11\bar20)_{\beta'_1}$ and $(c)$ $(11\bar20)_{\text{Mg}}\parallel(0001)_{\beta'_1}$.}
	\label{InterOR}
\end{figure}

To the light of the experimental observations in the previous section, the MS predictions of the interface energy were determined for the three most important interfaces between Mg and $\beta'_1$ precipitates. They stand for  $(0001)_{\text{Mg}}\parallel(0001)_{\beta'_1}$, $(0001)_{\text{Mg}}\parallel(11\bar20)_{\beta'_1}$ and $(11\bar20)_{\text{Mg}}\parallel(0001)_{\beta'_1}$ and the corresponding supercells are depicted in Fig. \ref{InterOR}. They were compared with the results obtained by DFT using the methodology detailed in \citep{RBP18} also based in supercell calculations.

Five different supercells were used to determine the interface energy for each interface orientation using MS. The interface area was constant in each supercell but the length $L$ was varied. All the supercells contained approximately the same number of atoms from both phases Mg and $\beta'_1$. 

Periodic boundary conditions were applied in all directions and the volume was allowed to expand only perpendicular to the interface. The energy of the system was minimized using the CG algorithm. Under these conditions, the excess of energy in the supercell, $\Delta E$, can be expressed as 

\begin{equation}
\Delta E=E_{tot}-(E_{\text{Mg}}+E_{\beta'_1})
\end{equation}

\noindent where $E_{tot}$  was the supercell energy obtained by MS and $E_{\text{Mg}}$ and $E_{\beta'_1}$ stand for the energy of the isolated Mg and $\beta'_1$ phases in the supercell.

The excess of energy can be attributed to two different sources according to

\begin{equation}\label{Int_MS2}
\frac{\Delta E}{2S}=E_{int}+\frac{1}{2}E_{el}L
\end{equation}

\noindent where $E_{int}$ stands for the interface energy per unit of area, $E_{el}$ the elastic energy per unit volume due to the elastic mismatch between Mg and $\beta'_1$, $L$  the supercell length perpendicular to the interface and $2S$  the interface area (two interfaces appear because of the periodic boundary conditions). Hence, the interface energy was given by the extrapolated value at $L$ = 0 of the straight line that relates $\Delta E/2S$ with $L$, which was determined from the calculations with different supercell lengths $L$ for each interface. 

The results obtained with MS (using the modified MEAM potential) and DFT for the different interfaces are shown  in Table \ref{ResultsOR}. The MS results overestimated slightly the interface energies  but the ranking of the different interfaces from the viewpoint of energies was correct, namely the $(0001)_{\text{Mg}}\parallel(0001)_{\beta'_1}$ interface (that never appears experimentally) has the highest interface energy while the two other interfaces have much lower interface energy, in agreement with the experimental observations \citep{SRS10}.

\begin{table}[H]
 \begin{center}
    \caption{Interface energies (mJ/m$^2$) of the different OR between Mg and $\beta'_1$-MgZn$_{2}$.} 
    \label{ResultsOR}
    \begin{tabular}{l|c|c|c} \hline \hline
     & $(0001)_{\text{Mg}}\parallel(0001)_{\beta'_1}$ &  $(0001)_{\text{Mg}}\parallel(11\bar20)_{\beta'_1}$ & $(11\bar20)_{\text{Mg}}\parallel(0001)_{\beta'_1}$ \\
      \hline 
		MS & 543 & 417 & 369 \\
		DFT &  483  & 332 & 202\\
     \hline \hline 
    \end{tabular}
  \end{center}
\end{table}

\section{Atomistic simulation methodology} \label{Methodology}

\subsection{Matrix/precipitate model}\label{IntModels}

The atomic positions of the different chemical species are known in the case of coherent interfaces with a given crystallographic orientation. This is not, however, the situation for semicoherent or incoherent interfaces. Building minimum energy atomistic interfaces  for a given crystallographic orientation  is important because it will influence the interactions mechanisms between the dislocation and the precipitate. The interface optimization methodology presented in \cite{EMP19} was used to build the atomistic Mg/$\beta'_1$ model.

According to the experimental evidence, ${\beta'_1}$ are long rods (Fig. \ref{precipitates}) and the top and bottom surfaces of the precipitate correspond to the orientation relationship  $(0001)_{\text{Mg}}\parallel(2\bar1\bar10)_{\beta'_1}$. There is not information, however, about the crystallographic orientation of the lateral surfaces of the rod and they were built from three main hypotheses. First,  it was assumed that lateral surfaces were faceted  and that two facets correspond to the $(2\bar1\bar10)_{\text{Mg}}\parallel(0001)_{\beta'_1}$ interface, that has the lowest interface energy according to the DFT calculations  (Table \ref{ResultsOR}). Several other interface possibilities were explored between the prismatic I (\{11$\bar 2$0\}$_{\text Mg}$) and prismatic II (\{1$\bar 1$00\}$_{\text Mg}$) planes of Mg and basal (\{0001 \}$_{\beta'_1}$), pyramidal ({1$\bar1$01\}$_{\\beta'_1}$), prismatic I (\{11$\bar 2$0\}$_{\text Mg}$ ), twin I (\{1$\bar 1$02 \}$_{\beta'_1}$) and twin II (\{1$\bar 1$03\}$_{\beta'_1}$)  of the $\beta'_1$ precipitate (Fig. \ref{Planes}). The energies of the different interfaces created from these planes were very similar (although always much higher than  the one of $(2\bar1\bar10)_{\text{Mg}}\parallel(0001)_{\beta'_1}$) and it was likely that several of them could co-exist, leading to a faceted lateral interface. 

\begin{figure}
	\centering
	\includegraphics[width=0.8\textwidth]{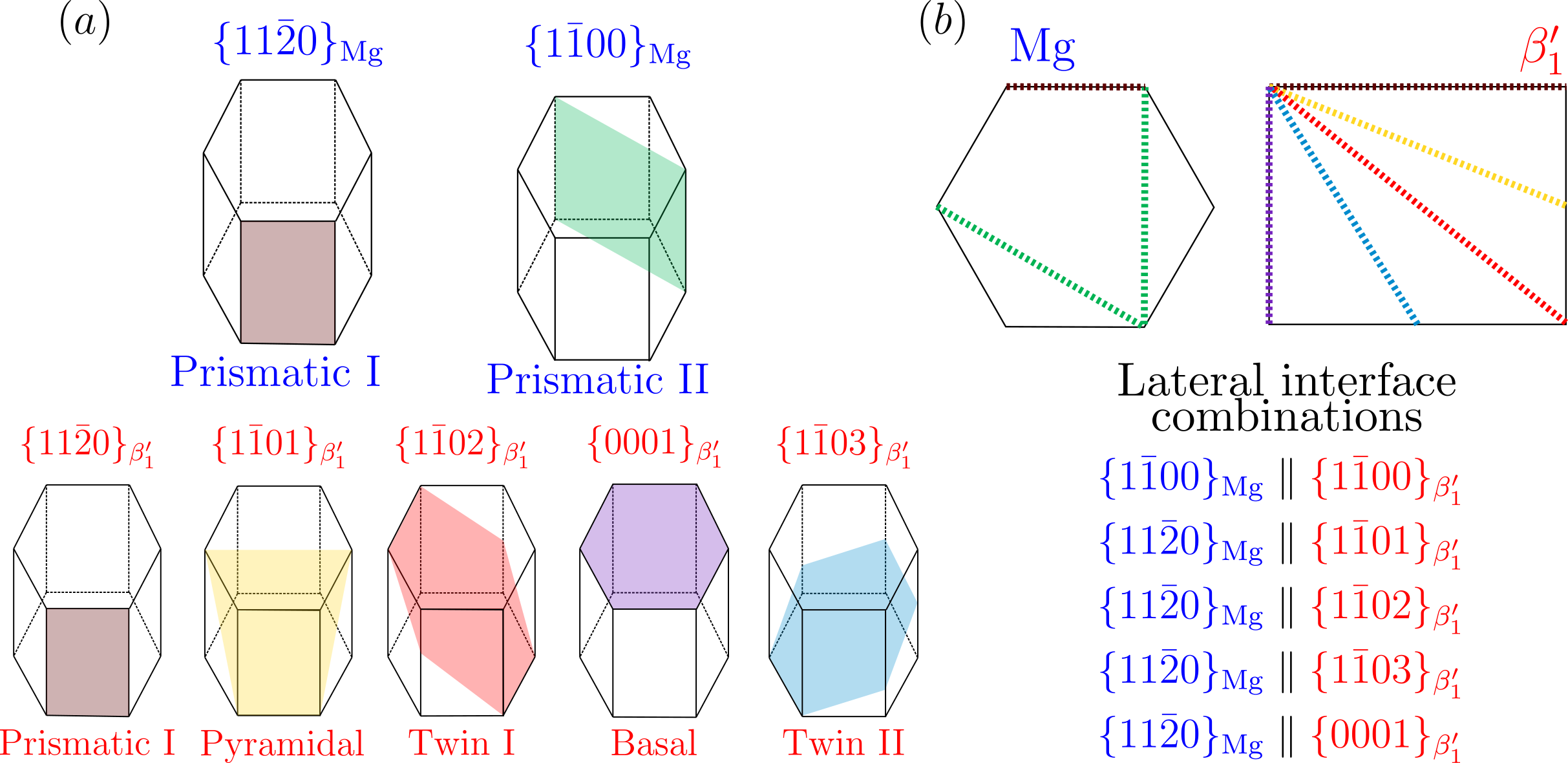}
	\caption{$(a)$ Possible families of planes that may conform the lateral faceted interface between Mg and $\beta'_1$ precipitate in addition of the $(2\bar1\bar10)_{\text{Mg}}\parallel(0001)_{\beta'_1}$ interface, which has the lowest interface energy. $(b)$ Top view (parallel to the $c$ axis of Mg) of the relative orientation between the lattices of Mg and $\beta'_1$ precipitate. Dashed lines represent the possible coincident planes in both phases to form lateral interfaces. The colour of each line stands for the corresponding colour of the plane in $(a)$.}
	\label{Planes}
\end{figure}

\color{black}Two $\beta'_1$ precipitates with initial faceted circular cross-sections of 3 and 10 nm in diameter were selected.\color{black} In order to create an atomistic model of Mg-$\beta'_1$ with the lowest lateral interface energy,  eleven faceted models of $\beta'_1$ precipitates were built up and inserted in a small Mg domain with dimensions 21.4 $\times$ 22.7 $\times$ 7 nm$^3$ along the $x$, $y$ and $z$ axes, respectively. These axes correspond to the crystallographic orientations $x=[1\bar{2}10]$, $y=[10\bar{1}0]$ and $z=[0001]$ of the Mg lattice. Different lateral interface combinations were selected from those in Fig. \ref{Planes}$(b)$ assuming that two facets have to be made by the $(2\bar1\bar10)_{\text{Mg}}\parallel(0001)_{\beta'_1}$ interface and that the precipitate cross-section has to be symmetric with respect to the $[10\bar10]_{\text{Mg}}\parallel[10\bar10]_{\beta_1'}$ and $[1\bar210]_{\text{Mg}}\parallel[0001]_{\beta_1'}$ directions (Fig. \ref{Interface}).

\begin{figure}
	\centering
	\includegraphics[width=0.9\textwidth]{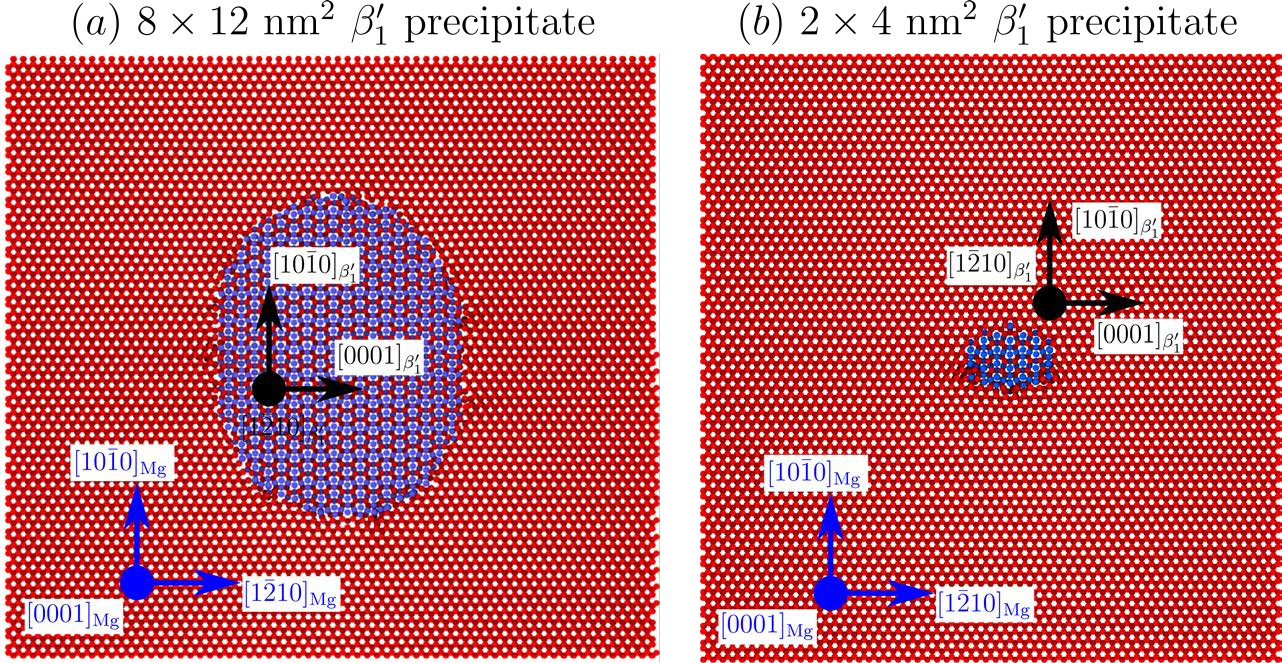}
	\caption{\color{black}Cross-section along the Mg basal plane  of the atomistic model showing the faceted interface with the lowest interface energy. $(a)$ 8 x 12 nm$^2$ $\beta'_1$ precipitate. $(b)$ 2 x 4 nm$^2$ $\beta'_1$ precipitate.  Both precipitates were surrounded by the combination of the   $\{2 1 \bar 1 0 \}_{\text{Mg}} || (0001)_{\beta'_1}$, $\{2 1 \bar 1 0 \}_{\text{Mg}} || \{1 \bar 1 0 3 \}_{\beta'_1}$ and  $\{2 1 \bar 1 0 \}_{\text{Mg}} || \{1 0 \bar 1 0 \}_{\beta'_1}$  interfaces. Although the elliptical cross-sections of the large and small precipitates have the largest and shortest axes in different orientations, it should be noted that lateral interfaces are the same in both precipitates. \color{black}}
	\label{Interface}
\end{figure}

Periodic boundary conditions were applied in all directions, accounting for an infinite rod precipitate embedded in the Mg matrix. The matrix and precipitate atoms were overlapped  and the Mg atoms within the precipitate were removed. However, matrix and precipitate atoms were still overlapped at the interface. Thus, a cut-off radius (in the range 0.25 nm to 0.35 nm) was defined and all the atoms belonging to the Mg matrix within the cut-off radius of an atom of the precipitate were deleted. The atomistic models obtained with different cut-off radii were minimized using the conjugate gradient. The energy minimization was initially carried out at constant volume and subsequently at zero stress. The excess of energy, $E_{ex}$, due to the lateral interface was calculated according to
 
\begin{equation}
E_{ex} = E_{tot}-(E_c^{\text{Mg}}n^\text{Mg}+E_c^{\beta'_1}n^{\beta'_1})
\label{E_excess}
\end{equation}

\noindent where $E_{tot}$ is the total potential energy of the system, and $E_c$ and $n$ stand for the cohesive energies and number of atoms, respectively, of each phase. 

The lateral interface with the lowest $E_{ex}$ was formed by the combination of the  interfaces  $\{2 1 \bar 1 0 \}_{\text{Mg}} || (0001)_{\beta'_1}$, $\{2 1 \bar 1 0 \}_{\text{Mg}} || \{1 \bar 1 0 3 \}_{\beta'_1}$ and  $\{2 1 \bar 1 0 \}_{\text{Mg}} || \{1 0 \bar 1 0 \}_{\beta'_1}$ and it was attained when the cut-off radius was 0.29 nm. 
\color{black} The final of the cross-section of the precipitates was approximately elliptical, as  shown in Fig. \ref{Interface}. Thus, the interface energy minimization process transformed the cross-section of the precipitates from circular into elliptical although the OR of the interfaces considered in the initial faceted shape did not change during the process. The length of each facet did change, however, transforming the circle into an ellipse. The maximum axis of the ellipse was parallel to the $[10 \bar10]_{\text{Mg}}$ in the case of the large precipitate (Fig. \ref{Interface}a) and parallel to the $[1\bar210]_{\text{Mg}}$ in the case of the small precipitate (Fig. \ref{Interface}b). \color{black} 

\subsection{Atomistic simulations}

An atomistic domain of 40.1 $\times$ 35 $\times 40.5$ nm$^{3}$ along the $x$, $y$ and $z$ axes, respectively, was used to analyse the dislocation/precipitate interactions. The axes were aligned along the directions $x=[1\bar{2}10]$, $y=[10\bar{1}0]$ and $z=[0001]$ of the Mg lattice. Periodic boundary conditions were applied along the $x$ and $y$ directions, while the boundaries perpendicular to the $z$ direction were free. This model stands for the periodic array of dislocations and precipitates presented by \cite{OB03}. The shape of the domain was selected to minimize the image stresses that appear as a result of the bowing of the dislocation during the simulations \citep{SC15}. The precipitate with the minimum lateral interface energy was created and inserted into the Mg matrix using the procedure presented above. An edge basal dislocation was also introduced in the domain by inserting a semi-plane of atoms into the model and applying the  displacement field corresponding to a dislocation in a anisotropic media \citep{atomsk}. The Burgers vector of the perfect basal dislocation was parallel to the $x$ axis, while the dislocation line was parallel to the $y$ axis. The schematic of the dislocation and the precipitate within the domain are depicted in Fig. \ref{AtomDom}$(a)$. The energy of the whole domain was minimized via CG algorithm, first at constant volume and afterwards at zero Virial stress. As a result, the perfect basal dislocation was split into two Shockley partials. Two different $\beta'_1$ precipitates were used in the simulations. Both precipitates have the same length of 20 nm parallel to the $c$ axis of the Mg lattice but have different cross-sections with minimum and maximum dimensions of 4 x 2 nm$^2$ and 12 x 8 nm$^2$. The latter dimensions are similar to the experimental ones (Fig. \ref{precipitates}).

\begin{figure}[!]
	\centering
	\includegraphics[width=\textwidth]{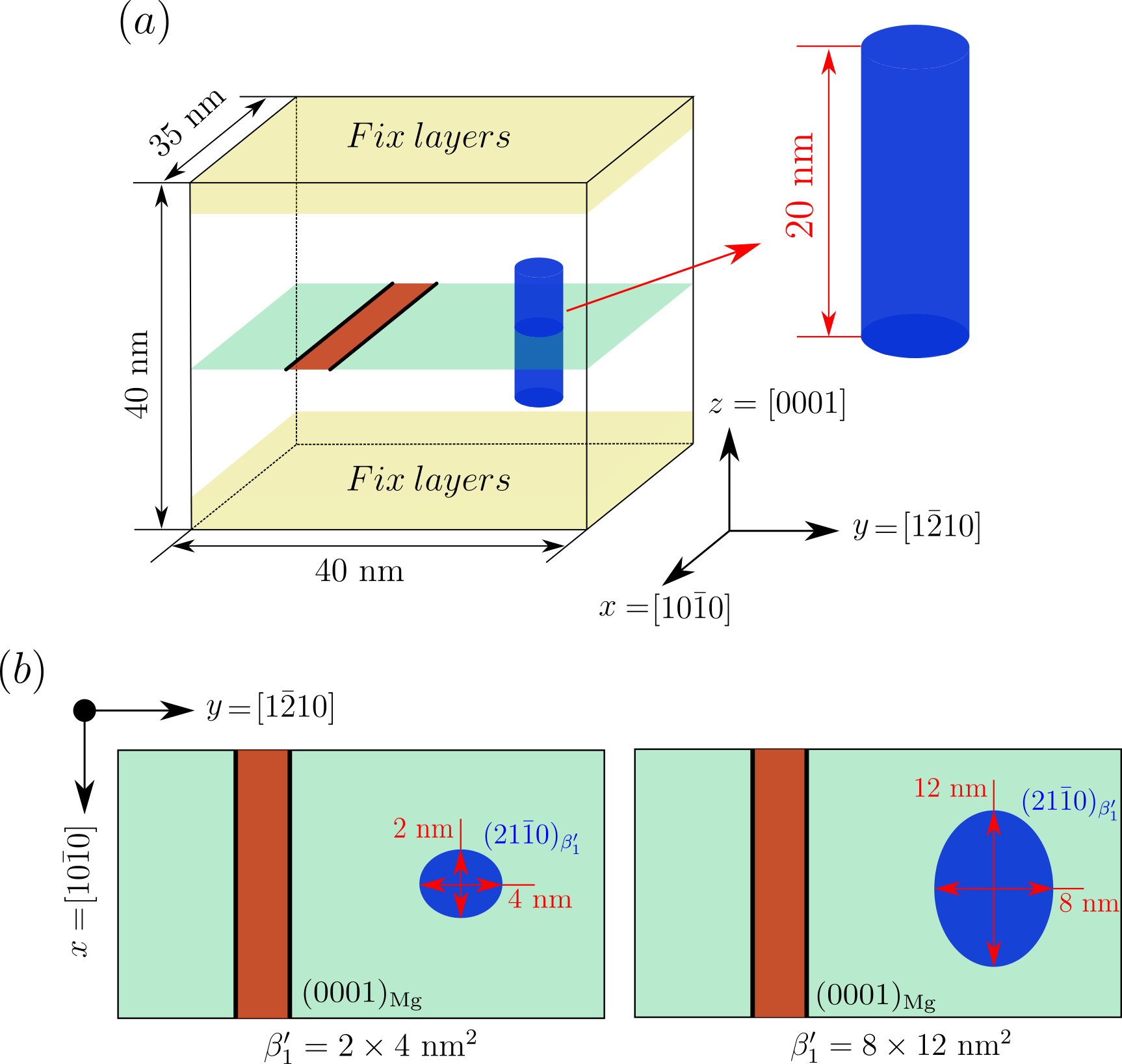}
	\caption{$(a)$ Representation of the atomistic domain used in MS and MD simulations. It contained a single basal dislocation and precipitate with periodic boundary conditions along the $x$ and $y$ axes. This model stand for a perfect array of dislocations and precipitates. $(b)$ Schematic representation of the shape, size and orientation of the two precipitates used in the atomistic simulations. }
	\label{AtomDom}
\end{figure}

MS simulations were used in the domain with the small precipitate in order to examine the basal dislocation/$\beta'_1$ precipitate interactions in the athermal limit. Shear displacements of 0.025 nm were applied successively in three atomic layers on the upper part of the domain, while three layers of atoms at the bottom surface remained fixed, Fig. \ref{AtomDom}$(a)$. The discrete displacements in the upper region were applied along the slip direction ($x$ direction) and the overall energy of the system was minimized at constant volume after each displacement. The process was repeated until two  dislocations overcame the precipitate (due to the periodic boundary conditions, the dislocations leaving the domain by one surface enter into the domain through the opposite surface).

MD simulations were carried out at 10 K, 100 K, 300 K and 400 K to evaluate the effect of temperature on the mechanisms of the dislocation/precipitate interaction and the CRSS in small and large precipitates (Fig. \ref{AtomDom}$(b)$). The atomistic domain was initially stabilized at the simulation temperature using the NPT ensemble during 30 ps while the volume was allowed to expand, relaxing the normal stresses. Afterwards, the NVT ensemble was applied and the atoms in the three top layers were displaced parallel to the slip plane to impose a shear strain rate of $\dot{\gamma}$ = 1.3 $\times$ 10$^8$ s$^{-1}$, while the atoms in the three bottom layers remained fixed. A timestep of 0.001 ps was used in the MD simulations and the stress, total energy and atomic position data were stored every 1 ps. The effect of the precipitate spacing was evaluated by increasing the length of the atomistic domain along the dislocation line ($x$ axis). The length of the domain was $15$ nm and $35$ nm for the small precipitate and 25, 35, 45 and 55 nm for the large one. All calculations were carried out using the open-source parallel molecular dynamics code LAMMPS \citep{lammps}. The results were visualized and analysed by means of  code OVITO \citep{ovt}. 

\section{Results and discussion} \label{Results}

\subsection{CRSS at 0 K}

MS simulations were carried out in the domain with the small precipitate (cross-section 2 x 4 nm$^2$) to assess the mechanisms of dislocation/precipitate interaction in the athermal limit. The shear stress and stored energy are plotted as function of the applied shear strain in Figs. \ref{MS_data}$(a)$ and $(b)$, respectively. The most relevant points of the interaction are labelled from $(i)$ to $(vi)$ in  Fig. \ref{MS_data}$(a)$ and the corresponding atomistic representation of the dislocation partials and of the precipitate atoms in the slip plane (coloured according to the shear strain) are  illustrated in Fig. \ref{MS_atom}. 
 
\begin{figure}[h]
	\centering
	\includegraphics[width=\textwidth]{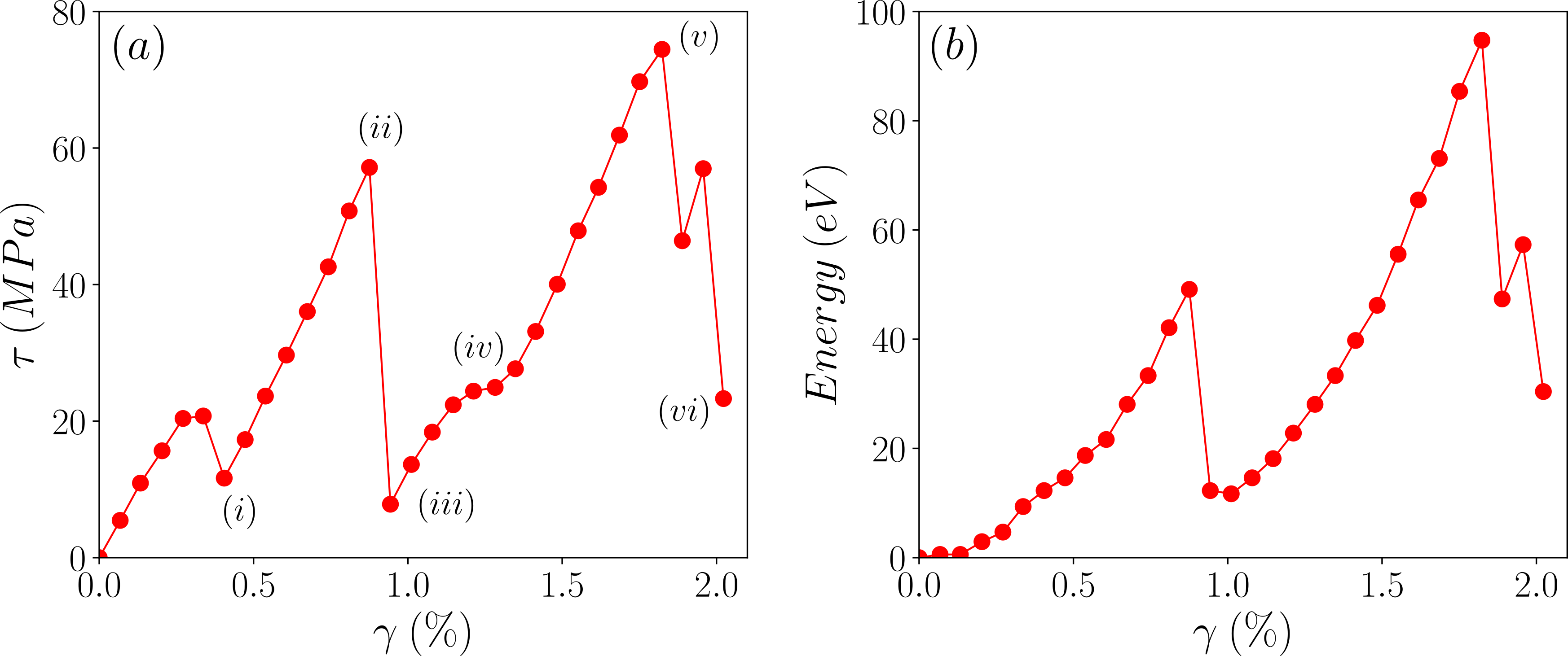}
	\caption{(a) Shear stress \textit{vs.} shear strain curve corresponding to the interaction of a basal dislocation with a $\beta_1'$ precipitate according to MS simulations. The atomic representation at the points labelled from $(i)$ to $(vi)$ can be found in Fig \ref{MS_atom}. (b) Stored energy \textit{vs.} shear strain curve according to MS simulations.}
	\label{MS_data}
\end{figure}

The stress necessary to move the dislocation increases initially with the applied strain until the applied strain reaches $\approx$ 0.3\% and the leading partial dislocation is attracted towards the precipitate, giving rise to the first local minimum in the shear stress-strain curve  (Fig. \ref{MS_data}a $(i)$) when the leading partial dislocation reaches the precipitate (Fig. \ref{MS_atom}$(i)$ and the matrix-precipitate interface is locally rearranged. Further deformation leads to a increase in the shear stress, which is associated with the formation of two Orowan loops around the precipitate by the partial dislocations. The precipitate is overcome by the dislocation at an applied shear strain of $\approx$ 0.9\% (Fig. \ref{MS_data}a $(ii)$) and both partial dislocations are recombined within the precipitate although the dislocation loop was not able to penetrate further into the precipitate (Fig. \ref{MS_atom}$(iii)$). Further deformation leads to the propagation of the second dislocation and the leading partial touches first the precipitate, Fig. \ref{MS_atom}$(iv)$,  although the strong attraction between them has disappeared because of the dislocation loop within the precipitate. The second dislocation also overcomes the precipitate by the formation of an Orowan loop, Fig. \ref{MS_atom}$(v)$, which pushes the first loop to shear the $\beta'_1$, as indicated by the large shear strains within the precipitate (Fig. \ref{MS_atom}$(vi)$) and is accompanied by a large reduction in the stress and energy stored (Fig. \ref{MS_data}).

\begin{figure}[h]
	\centering
	\includegraphics[width=0.7\textwidth]{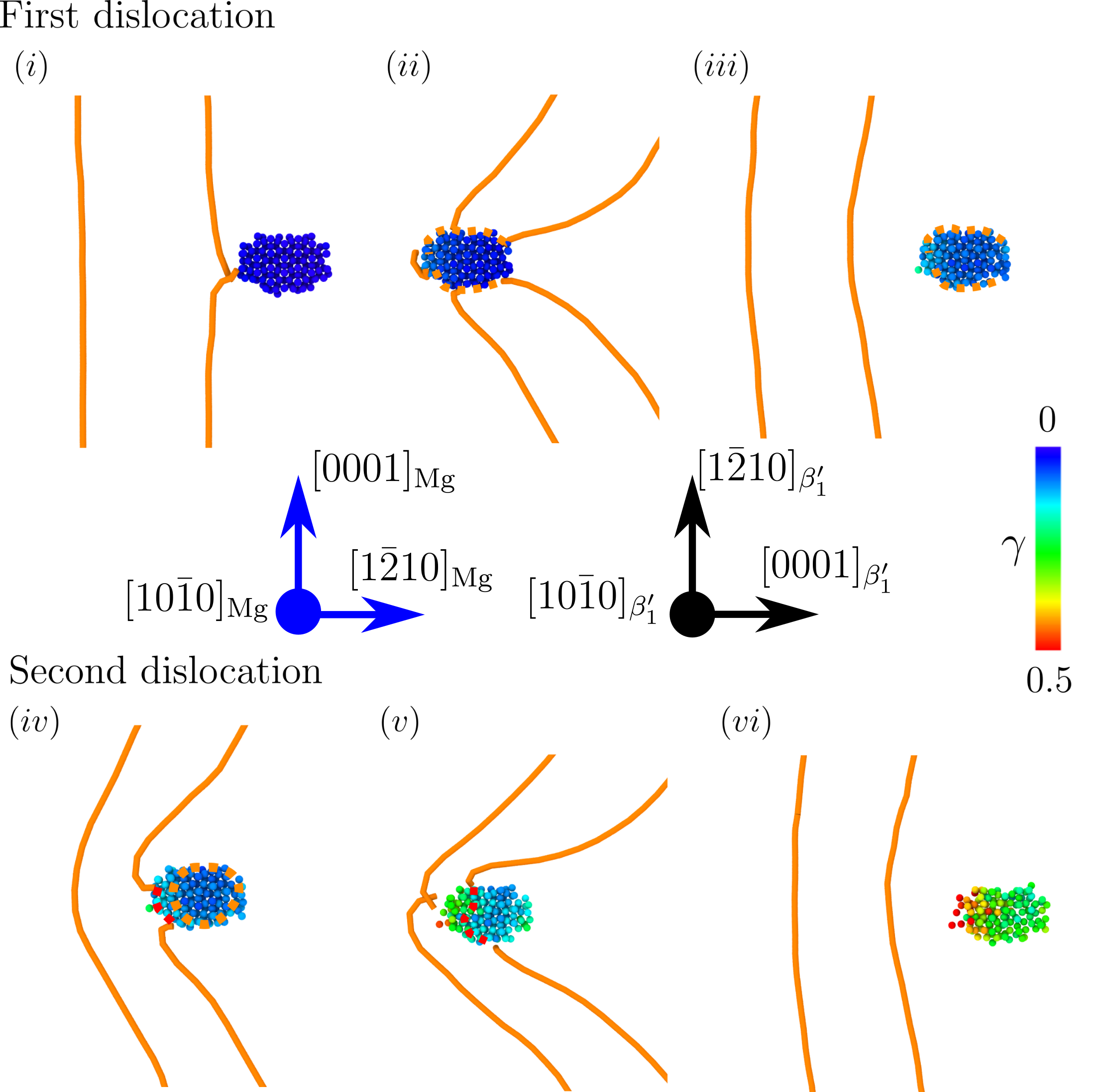}
	\caption{Atomistic representation across the slip plane of the interaction of a basal dislocation with a $\beta_1'$ precipitate according to MS simulations. The label in each picture is referred to the point with the same label in the shear stress-strain curve in Fig. \ref{MS_data}$(a)$. The orange lines indicate the leading and trailing dislocation partials. The interaction of the two Shockley partials corresponding to the first full dislocation with the precipitate is depicted from $(i)$ to $(iii)$, while $(iv)$ to $(vi)$ stand for the interaction of the  Shockley partials of the second full dislocation with the precipitate. The colour in the precipitate atoms denotes the atomic shear strain.}
	\label{MS_atom}
\end{figure}

\subsection{CRSS at finite temperatures}

MD simulations were carried out in the domain with the small (cross-section 2 x 4 nm$^2$) and large (cross-section 8 x 12 nm$^2$) precipitate at \color{black} 10 K, 100 K, 300 K and 400 K \color{black} to assess the effect of temperature on the interaction mechanisms.

The shear stress - shear strain curves at different temperatures corresponding to the small precipitate are plotted in Fig. \ref{MD_2}$(a)$ and $(b)$ for two different values of the precipitate spacing $\lambda$ = 15 nm and 35 nm, respectively. They were obtained with different values of the width of the simulation box along the $x$ axis (Fig. \ref{AtomDom}$(a)$). The stress increases linearly with the applied shear strain at the beginning of the deformation in all cases. Nevertheless, the curves begin to diverge as the leading dislocation approaches the precipitate in the case of $\lambda$ = 15 nm and, in fact, the stress necessary to overcome the precipitate (indicated by the first maximum in  Fig. \ref{AtomDom}$(a)$) increases as the temperature decreased. Moreover, the first dislocation loop is able to shear the precipitate in the simulations at \color{black} 300 K and 400 K \color{black} but not in the simulations at \color{black} 10 K and 100 K. \color{black} The dislocation loop is able to penetrate the precipitate (as in the static simulations) but it cannot to shear it until the second dislocation loop pushes the first one through the precipitate. Higher stresses are necessary to move the second dislocation near the precipitate in these cases because of the repulsion with the first dislocation loop inside the precipitate but, once the precipitate has been sheared, the shear stress necessary to move more dislocations decreases. Nevertheless,  the CRSS (understood as the absolute maximum in the shear stress - strain curve) decreases as the temperature increases in this case. Similar results were obtained when the distance between precipitates was $\lambda$ = 35 nm (Fig. \ref{AtomDom}$(b)$). The precipitate is sheared by the first dislocation at \color{black} 300 K and 400 K \color{black}, by the second dislocation at \color{black} 100 K \color{black} and by the third dislocation at \color{black} 10 K \color{black}. Thus, the CRSS also decreases with temperature although the dependence is smaller than when $\lambda$ = 15 nm. 

\begin{figure}[h]
	\centering
	\includegraphics[width=\textwidth]{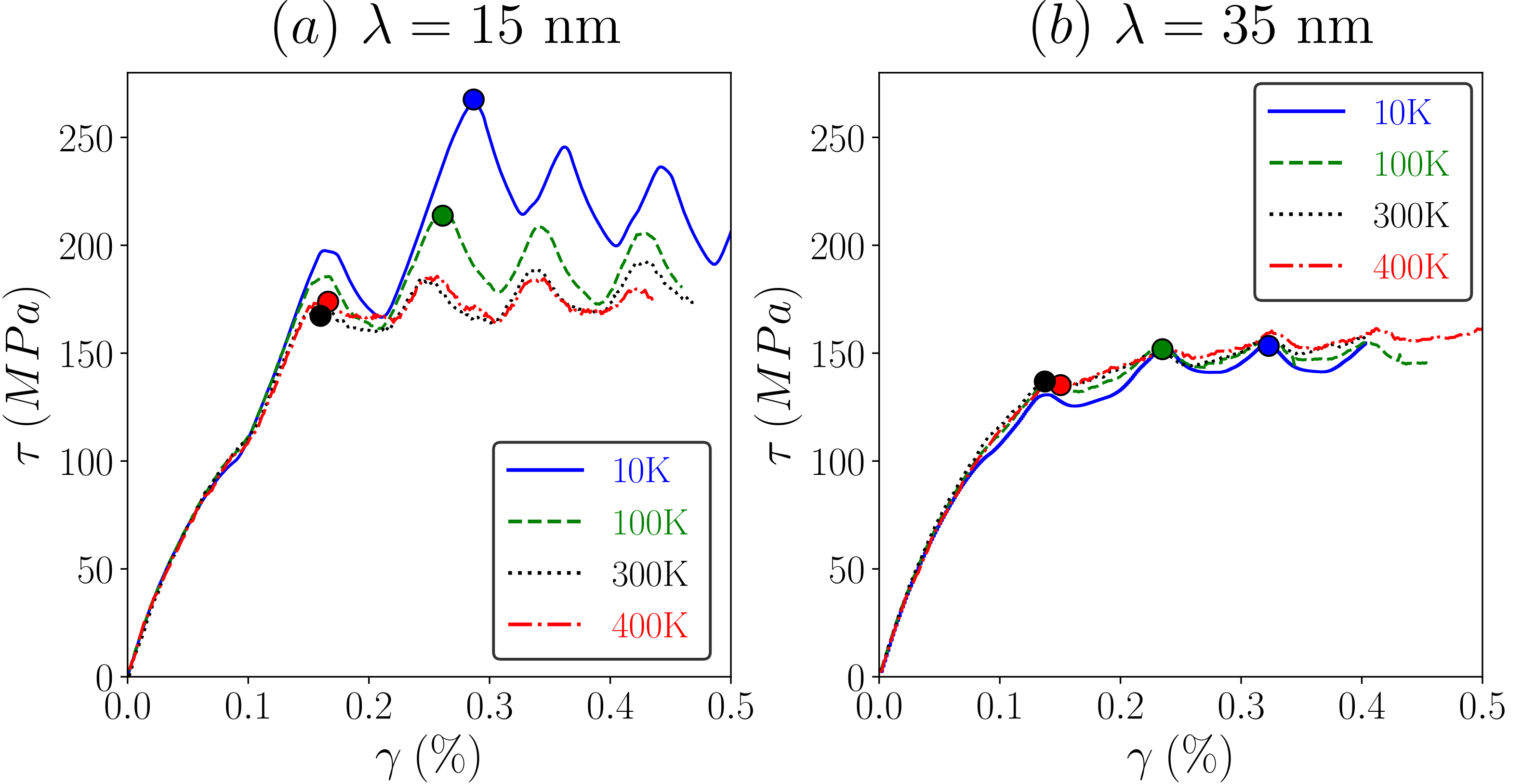}
	\caption{Shear stress-strain curves obtained from MD calculations for the small $\beta_1'$ precipitate as a function of temperature.  $(a)$ The distance between precipitates along the dislocation line was $\lambda$ = 15 nm. $(b)$ $\lambda$ = 33 nm. The dots on the curves denote the instant when the precipitate was sheared.}
	\label{MD_2}
\end{figure}

The shear stress-strain curves obtained by MD of the large precipitate are plotted as  a function of temperature \color{black}(from 10 K to 400 K) \color{black} for precipitate spacing of $\lambda$ = 25, 35, 45 and 55 nm in Figs. \ref{MD_12}$(a)$ to $(d)$, respectively. The mechanisms of dislocation/precipitation interaction are equivalent to those found in MS and MD of small precipitates. Dislocations overcome the precipitate by the formation of an Orowan loop that enters the precipitate. The precipitate is eventually sheared after several dislocations overcome the precipitate and the number of dislocations to shear the precipitate decreases with temperature and is independent of the distance between precipitates. In all these curves, the shear stress increases initially more or less linearly with the shear strain as the dislocation approaches the precipitate. The first minimum in the curve is triggered by the attraction of the leading partial towards the precipitate and the first Orowan loop to overcome the precipitate is given by the second maximum in the  shear stress-strain curve. Thus 7 Orowan loops are required to shear this large precipitate at \color{black}10 K, 4 loops at 100 K and 3 loops at 300 K and 400 K \color{black} and the CRSS to overcome the precipitates decreases with temperature regardless of the distance between precipitates. 

\begin{figure}[!t]
	\centering
	\includegraphics[width=\textwidth]{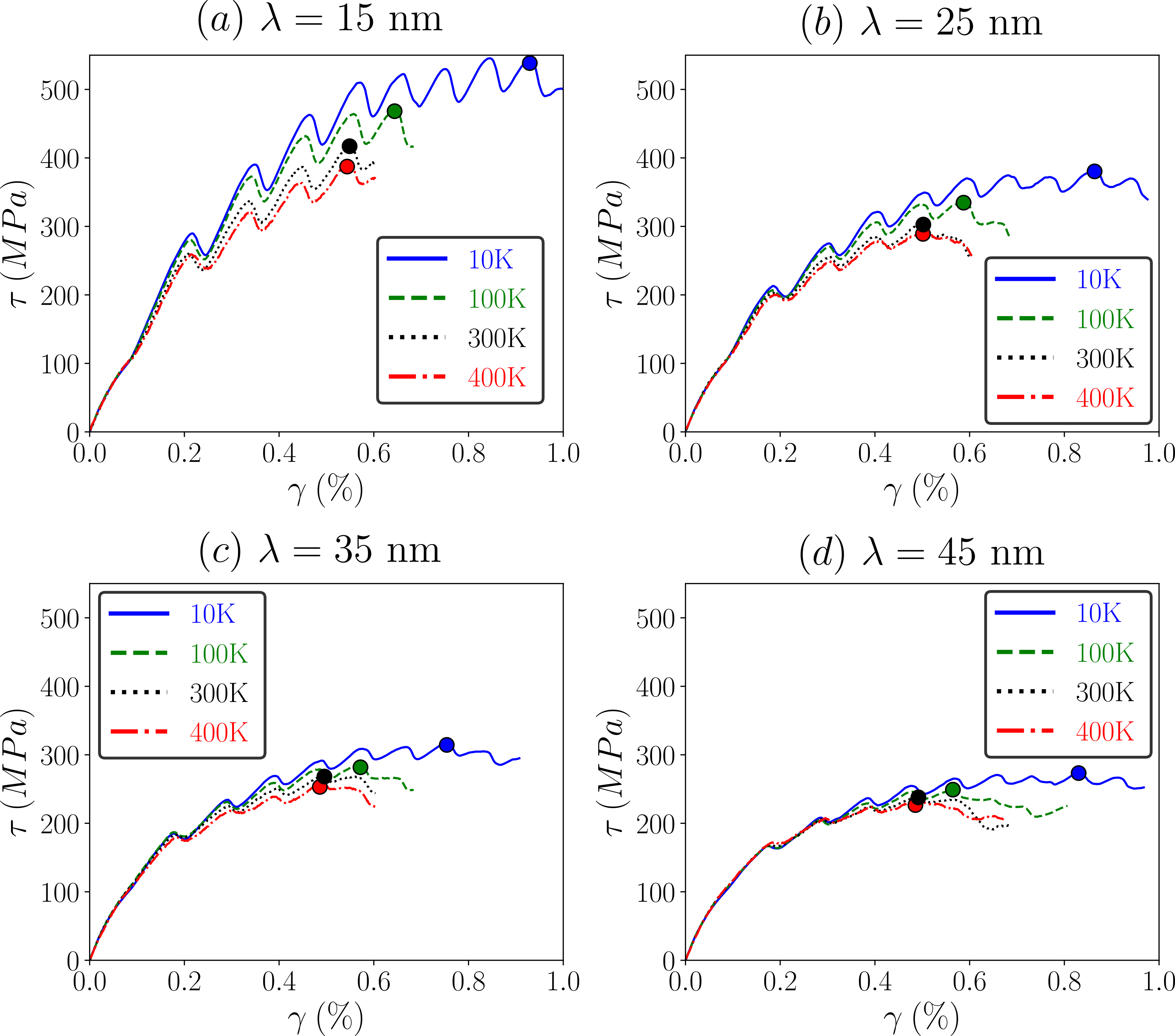}
	\caption{Shear stress-strain curves obtained from MD calculations for the large $\beta_1'$ precipitate as a function of temperature.  $(a)$ The distance between precipitates along the dislocation line was $\lambda$ = 15 nm. $(b)$ $\lambda$ = 25 nm. $(c)$ $\lambda$ = 35 nm. $(d)$ $\lambda$ = 45 nm. The dots on the curves denote the instant when the precipitate was sheared.}
	\label{MD_12}
\end{figure}

The overall effect of the precipitate spacing on the CRSS at different temperatures is plotted in Fig. \ref{CRSS} for $\beta_1'$ precipitates with small and large cross-section. Although the actual values of the CRSS are much higher than those observed experimentally \citep{CB69, WS15} --due to the huge strain rates associated with MD simulations--, the simulation results report that CRSS of precipitate-strengthened Mg-Zn alloy should decrease with temperature, in agreement with the experimental results \citep{CB69, AL19}. This dependence is due to the shearing mechanism that it is thermally activated while the by-pass of precipitates by the formation of Orowan loops is an athermal process. The effect of temperature is more noticeable in the large precipitates, while no differences in the CRSS between \color{black} 300 K and 400 K \color{black} were found in the case of the precipitates with small cross-section because they were always sheared by the first dislocation.

\begin{figure}[t]
	\centering
	\includegraphics[width=0.8\textwidth]{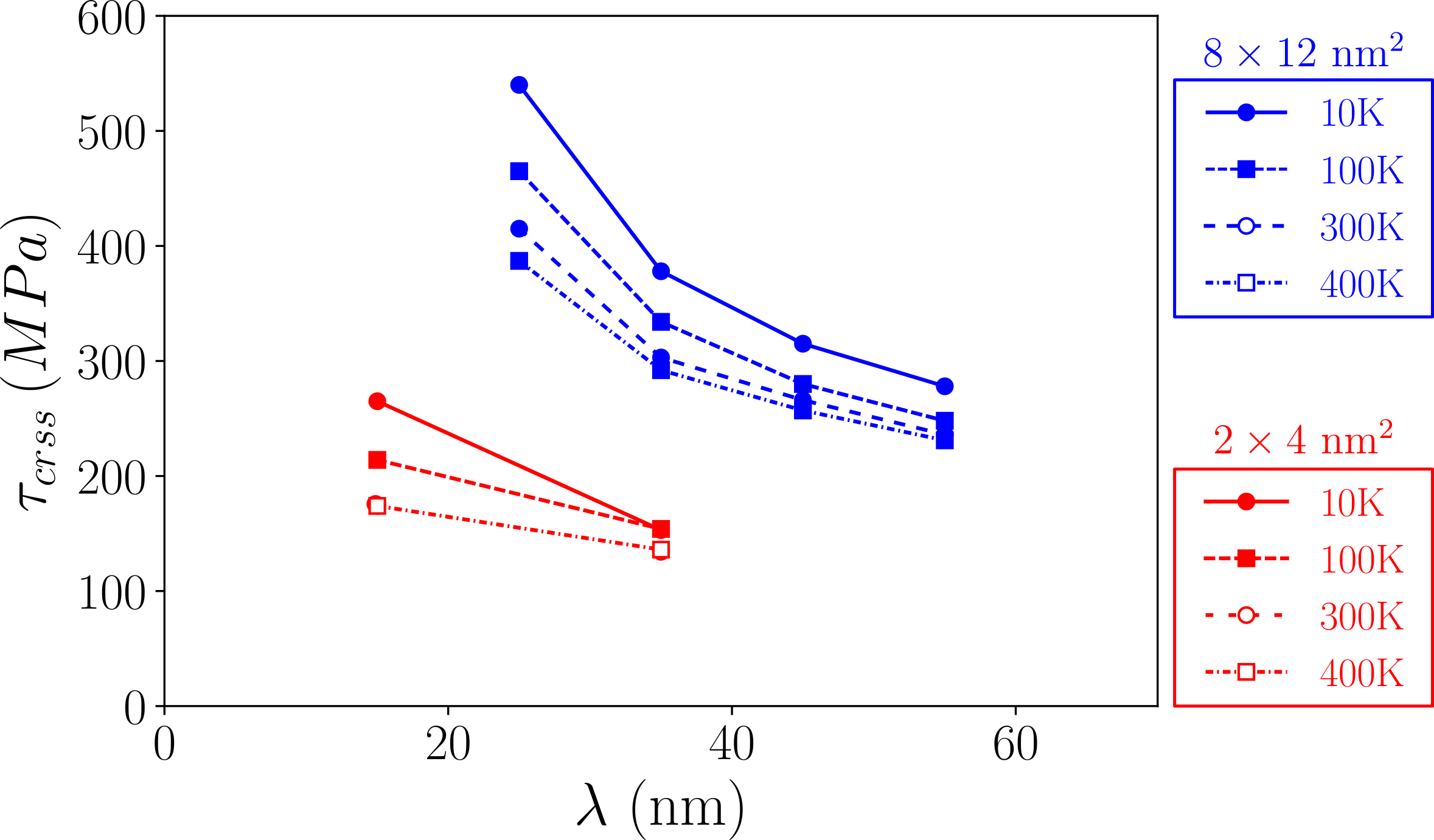}
	\caption{MD predictions of the CRSS for shearing of $\beta_1'$ precipitates by basal dislocations as a function of the precipitate spacing and temperature. Results for precipitates with small (2 x 4 nm$^2$) and large (8 x 12 nm$^2$) cross-section are shown. \color{black} In the case of small precipitate, $\tau_{crss}$ is the same at 300 K and 400 K. \color{black}}
	\label{CRSS}
\end{figure}

\subsection{Dislocation/precipitate interaction mechanisms}

The shear mechanism of the large $\beta_1'$ precipitate by basal dislocations at \color{black}300 K \color{black} is depicted in Figs. \ref{Atom_12nm}$(a)$ and $(b)$, which show the precipitate atoms in two different perspectives (lateral view and cross-section parallel to the slip plane) at different stages of the deformation. Different colours represent the shear strain parallel to the basal plane of the Mg matrix. The (i) lateral and cross-section images stand for the precipitate at the beginning of the simulation and the corresponding images at (ii), (iii) and (iv) show the precipitate deformation after the second, third and fourth dislocations have overcome the precipitate, respectively. The interaction of the dislocations with the precipitate led to the progressive localization of the deformation in the precipitate along a thin band parallel to the basal slip plane of Mg, Fig. 
\ref{Atom_12nm}$(a)$ (ii) to (iv). The contour plot of the shear strain in the cross section also shows that the precipitate was more easily penetrated by the dislocations along the [10$\bar 1$0]$_{\beta'_1}$, Fig. 
\ref{Atom_12nm}$(a)$ (iii). Finally, the strain gradients and the order in the (21$\bar 1$0)$_{\beta'_1}$ prismatic plane of the precipitate disappeared after the fourth dislocation sheared the precipitate, Fig. \ref{Atom_12nm}$(a)$ (iv). The lateral cross-section along the center of the precipitate is shown in Fig. 
\ref{Atom_12nm}$(c)$ at this stage and the kink in the precipitate is in good agreement with the TEM observations in Fig. \ref{DPinteraction23}.

\begin{figure}[!]
	\centering
	\includegraphics[width=0.92\textwidth]{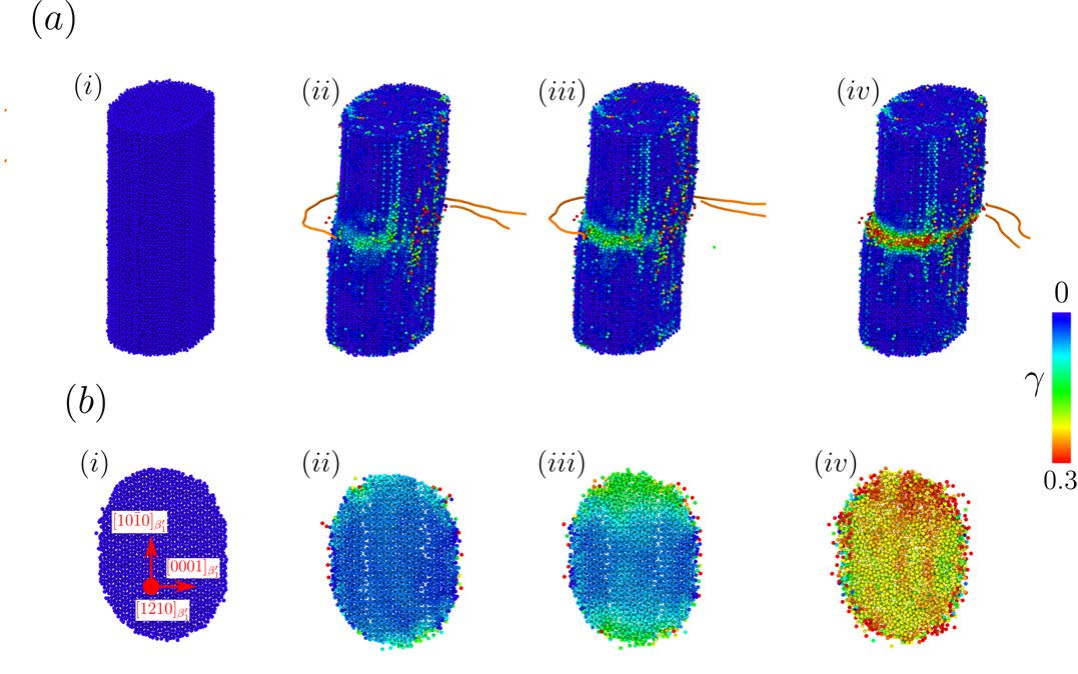}
	\includegraphics[width=0.65\textwidth]{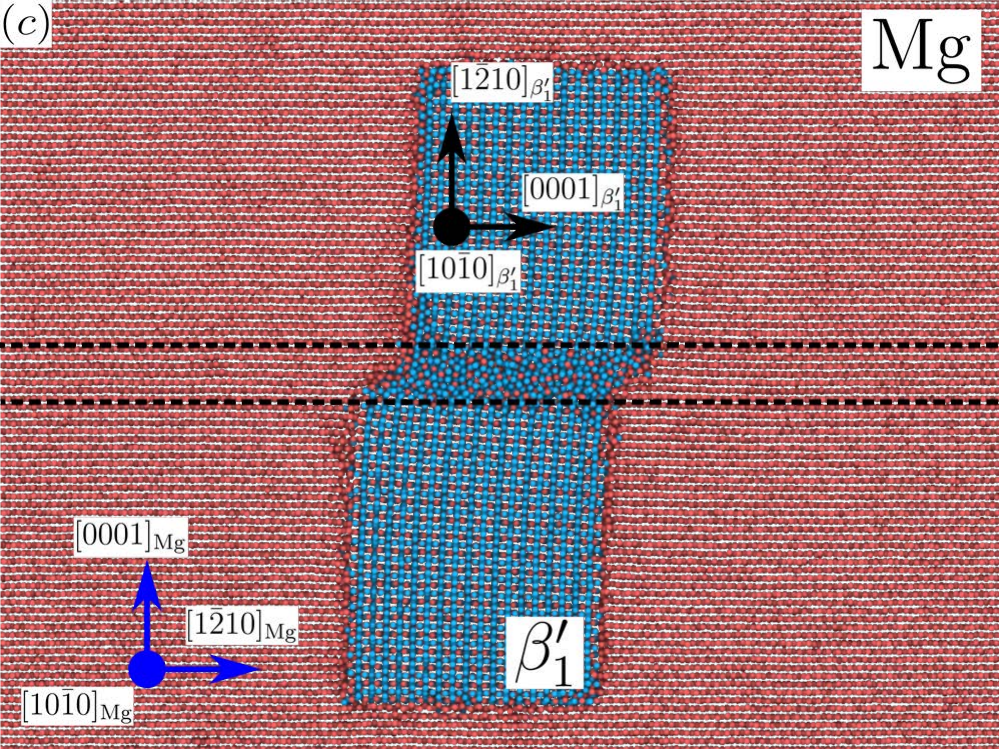}
	\caption{Atomic representation of the shearing of the large $\beta_1'$ precipitate by basal dislocations at \color{black}300 K. $(a)$ Lateral view. $(b)$ Cross-section parallel to the slip plane. Only the precipitate atoms and the partial dislocations (red lines) in the Mg matrix are shown. The precipitate atoms are coloured as a function of the shear strain parallel to the basal plane of the Mg matrix. The labels $(i)$, $(ii)$, $(iii)$, $(iv)$ correspond to the unstrained precipitate and after the second, third and fourth dislocations have overcome the precipitate, respectively. $(c)$ Lateral cross-section of the matrix and precipitate along a plane through the center of the precipitate after the fourth dislocation has sheared the precipitate. It should be noted that  final shape of the precipitate is in good agreement with the TEM images in Fig. \ref{DPinteraction23}. Mg atoms are red and Zn atoms are blue.}
	\label{Atom_12nm}
\end{figure}

In order to ascertain the actual mechanism of precipitate shearing, the positions of the atoms in the precipitate core were analysed during the dislocation/precipitate interaction and they are depicted in Fig. \ref{Mechanism}. The precipitate core before deformation presents  an ordered structure, Fig. \ref{Mechanism}$(a)$, in which prismatic planes of the precipitate (blue lines) are perpendicular to the basal plane of the Mg matrix. The interaction of the dislocations with the precipitate lead to the accumulation of elastic strain in the precipitate, which is shown by the rotation of the prismatic planes near the basal plane of Mg, Fig. \ref{Mechanism}$(b)$. Once the elastic energy stored in the precipitate reaches a critical value,  the order in the precipitate is lost, leading to the shearing of the precipitate and to the formation of an amorphous layer below and above the slip plane of the basal dislocations. The whole process is schematically shown in Fig. \ref{Mechanism} below the atomistic representations.  It is interesting to notice that, although the shearing of the precipitate does not take place along a crystallographic plane, the relative horizontal displacement between the upper and lower parts of the precipitate is around 0.85 nm, which is equivalent to the lattice parameter of the $\beta'_1$ precipitate along the $c$ axis.

\begin{figure}[t]
	\centering
	\includegraphics[width=\textwidth]{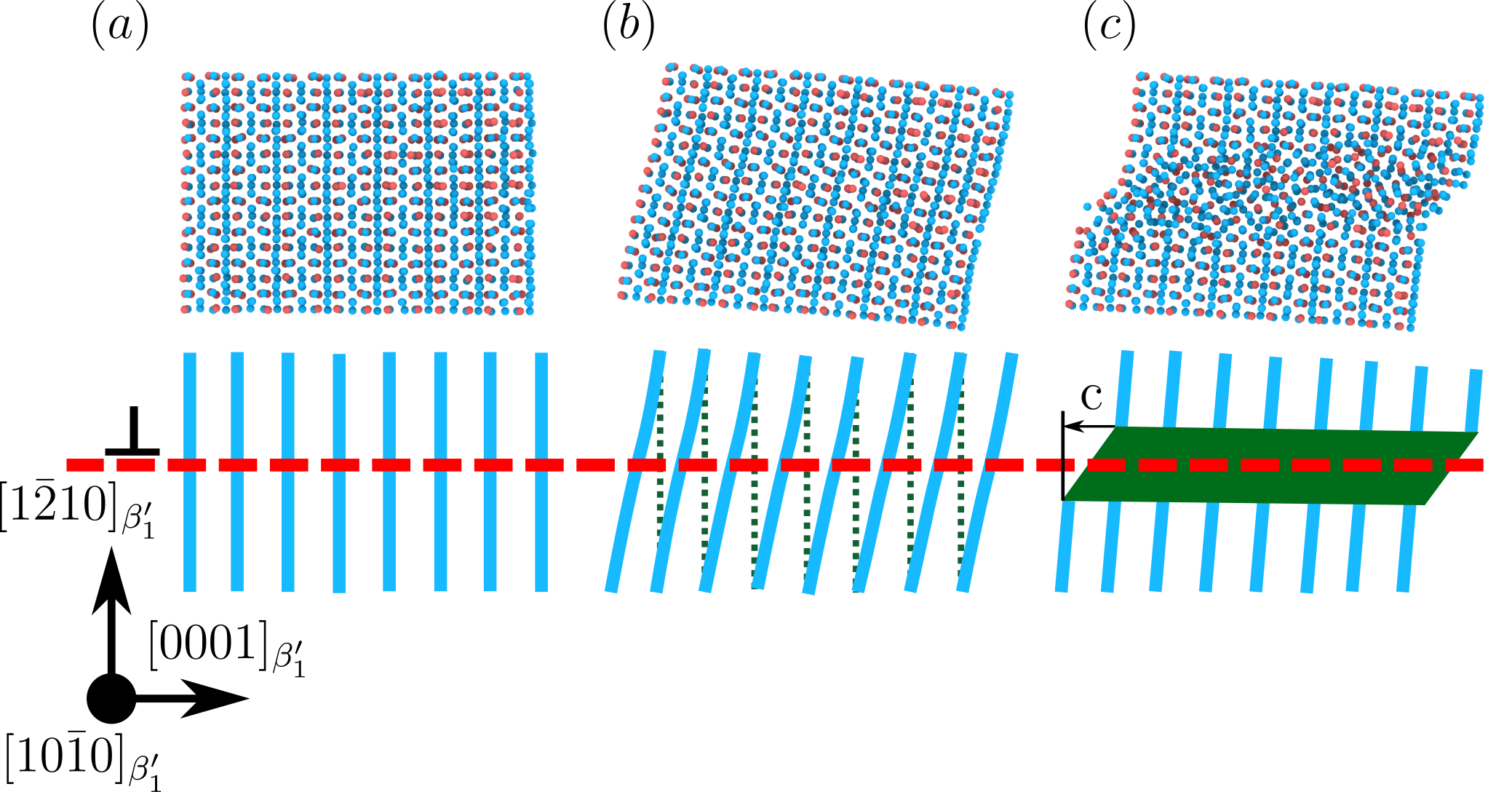}
	\caption{Atomistic representation of the precipitate core. $(a)$ Initial condition. $(b)$ At the critical instant previous to shear. $(c)$ After shearing the precipitate. Below each atomistic representation,  a schematic of the mechanism is depicted. The broken red line stands for the slip plane of the basal dislocation in Mg, the blue lines are the prismatic planes of the $\beta'_1$ precipitate and the green area is the disordered atomistic domain after shearing. Mg atoms are red and Zn atoms are blue.}
	\label{Mechanism}
\end{figure}

\color{black} In order to understand the mechanisms of dislocation/precipitate interaction, the atomic arrangement in the $(1\bar{2}10)$ plane of the precipitate, parallel to the basal slip plane in Mg, was analyzed. The radial distribution function (RDF) of Mg-Mg, Zn-Zn and Mg-Zn atom pairs was evaluated up to a distance of 1 nm in the central and right regions of the 12 x 8 nm$^2$ precipitate before any interaction with the dislocations (Fig. \ref{PlanesDef}a)  and after 5 (Fig. \ref{PlanesDef}b) and 7 dislocations (Fig. \ref{PlanesDef}c)  are accumulated at the interface. The RDF in the initial condition shows the peaks corresponding to the regular lattice of the precipitate. After 5 dislocations have pile-up at the interface (Fig. \ref{PlanesDef}b), the intensity of the peaks in the RDF (which indicate the probability of finding another atom at a given distance) has decreased, particularly in the right region near the precipitate/matrix interface. The accumulation of more dislocations at the interface (Fig. \ref{PlanesDef}c) led to further reductions in the intensity of the peaks, which have disappeared for pair distances $>$ 0.4 nm in the both the lateral and central regions of the precipitate. The evolution of the RDF is indicative of the progressive amorphization of the precipitate as a result of the stress concentrations associated with the accumulation of dislocations at the interface. Once the whole precipitate section has become amorphous, the arrival of new dislocations leads to the shearing of the amorphous layer without any increment in the CRSS.\color{black}

\begin{figure}[!t]
	\centering
	\includegraphics[width=\textwidth]{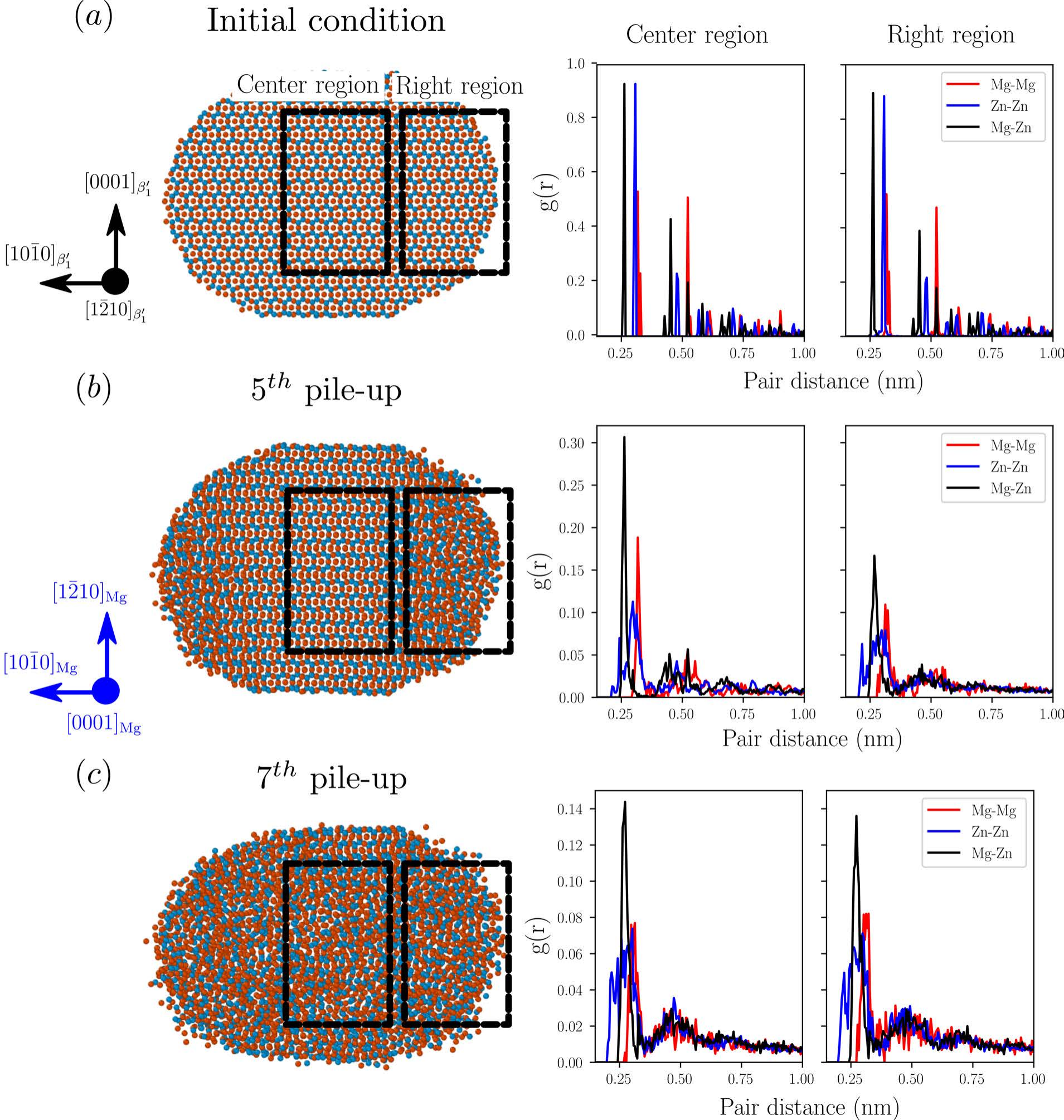}
	\caption{\color{black} Atomic arrangement in the $(1\bar{2}10)$ plane of the precipitate, parallel to the basal slip plane in Mg. $(a)$ Initial condition. $(b)$ After accumulation of 5 dislocations at the interface. $(c)$ After accumulation of 7 dislocations at the interface. The radial distribution function of Mg-Mg, Zn-Zn and Mg-Zn atom pairs in the central and right regions of the precipitate is depicted in each case. The long range order disappears gradually with the formation of dislocation pile-ups, starting from the right and left regions of the precipitate. \color{black}}
	\label{PlanesDef}
\end{figure}

\section{Conclusions} \label{Conclusion}

The interaction between Mg edge basal dislocations and rod-shaped $\beta_1'$-MgZn$_2$ precipitates has been examined by atomistic simulations using a modified interatomic potential which provided good estimations for the elastic constants and interfaces energies of the Mg matrix and the precipitates. The interfaces between the rod-shaped precipitates and the Mg matrix were carefully built taking into account the experimental information about the orientation relationship between the matrix and the precipitate to ensure that they were minimum energy interfaces. The top and bottom matrix/precipitate interfaces were   $(0001)_{\text{Mg}}\parallel(2\bar1\bar10)_{\beta'_1}$ while the lateral interfaces were formed by a combination of the  $\{2 1 \bar 1 0 \}_{\text{Mg}} || (0001)_{\beta'_1}$, $\{2 1 \bar 1 0 \}_{\text{Mg}} || \{1 \bar 1 0 3 \}_{\beta'_1}$ and  $\{2 1 \bar 1 0 \}_{\text{Mg}} || \{1 0 \bar 1 0 \}_{\beta'_1}$ interfaces.

The interaction mechanisms between edge basal dislocations and precipitates with small (2 x 4 nm$^2$) and large (8 x 12 nm$^2$) cross-section were studied using molecular static and molecular dynamics, the latter between \color{black} 10 K and 400 K \color{black}, and they were equivalent. It was found that the dislocations initially overcame the precipitate by the formation of an Orowan loop that penetrated in the precipitate. The precipitate was finally sheared after several Orowan loops were piled-up and the number of loops necessary to shear the precipitate decreased as precipitate cross-section decreased and the temperature increased but it was independent of the precipitate spacing. Thus, the molecular dynamics simulations indicated that the critical resolved shear stress to overcome the precipitates decreased with temperature in Mg-Zn alloys.

Precipitate shearing did not take place along well-defined crystallographic planes but it was triggered by the accumulation of the elastic energy in the precipitate which finally led to formation of an amorphous layer below and above the slip plane of the basal dislocations. The kink induced in the precipitate by this mechanism was in good agreement with transmission electron microscopy observations.

\section*{Acknowledgements}
This investigation was supported by the European Research Council under the European Union's Horizon 2020 research and innovation program (Advanced Grant VIRMETAL, grant agreement No. 669141). The computer resources and the technical assistance provided by the Centro de Supercomputaci\'on y Visualizaci\'on de Madrid (CeSViMa) are gratefully acknowledged. Additionally, the authors also acknowledge the computer resources at Picasso and the technical support provided by Barcelona Supercomputing Center (project QCM-2019-1-0004). RA also acknowledges the support from the Spanish Ministry of Science through the Juan de la Cierva program (FJCI-2016-29660).

\appendix 

\setcounter{table}{0}
\setcounter{figure}{0}
\setcounter{section}{0}
\renewcommand{\thetable}{A\arabic{table}}
\renewcommand{\thefigure}{A\arabic{figure}}
\renewcommand{\thesection}{A.\arabic{section}}

\section{Modification of MgAlZn 2NN-MEAM potential}
\label{AppMP}
The MEAM parameters for the pure metals (Mg, Al and Zn) and binary interactions (Mg-Al, Mg-Zn and Al-Zn) are shown in tables \ref{unary} and \ref{binary}, respectively. The parameters changed with respect to the MEAM potential in \cite{DD18} are in bold.

\begin{table}[H]
 \begin{center}
    \caption{MEAM potential parameters for the pure Mg, Al, and Zn interatomic potentials. The units of the cohesive energy ($E_c$) and of the lattice parameters ($a_{lat}$) are eV and $\text{\AA}$, respectively. The reference structures are HCP for Mg and Zn and FCC for Al} 
   \label{unary}
\resizebox{\textwidth}{!}{%
    \begin{tabular}{l|c|c|c|c|c|c|c|c|c|c|c|c|c|c|c|c} \hline \hline
      & $E_c$ & $a_{lat}$ & $\alpha$ & $A$ & $\beta^{(0)}$ & $\beta^{(1)}$ & $\beta^{(2)}$ & $\beta^{(3)}$ & $t^{(1)}$ & $t^{(2)}$ & $t^{(3)}$ & $C_{min}$ & $C_{max}$ & Attr & Rep & $\rho$  \\
     \hline
     		Mg & 1.51 & 3.19 & 5.608 & 0.52 & 2.00 & 1.30 & 1.30 & 1.00 & 5.55 & 3.00 & -7.40 & 0.49 & 2.90 & 0.00 & 0.00 & 1.00 \\
     		
     		Al & 3.36 & 4.05 & 4.69 & 1.16 & 3.20 & 2.60 & 6.00 & 2.60 & 3.05 & 0.51 & 7.75 & 0.49 & 2.90 & 0.00 & 0.00 & 1.175 \\
     		
     		Zn & 1.325 & 2.785 & 6.95 & 0.70 & 2.00 & 1.30 & 1.30 & 6.50 & 25.0 & 17.3 & 51.5 & 1.20 & 2.50 & 0.10 & 0.10 & 0.84 \\
    \hline \hline
    \end{tabular}
    }%
  \end{center}
\end{table}

\begin{table}[H]
 \begin{center}
    \caption{MEAM potential parameters for the binary Mg-Al, Mg-Zn, and Al-Zn interatomic interactions. The parameters $R_e$ and $E_c$ are expressed  in $\textbf{\AA}$ and $eV$, respectively.} 
   \label{binary}
\resizebox{\textwidth}{!}{%
    \begin{tabular}{l|c|c|c|c|c|c|c|c|c|c|c|c|c} \hline \hline
     L$_{12}$ & $R_e$ & $E_c$ & $\alpha_{A-B}$ & $C^{\text{A-A-B}}_{max}$ & $C^{\text{B-B-A}}_{max}$  & $C^{\text{A-B-A}}_{max}$  & $C^{\text{A-B-B}}_{max}$ & $C^{\text{A-A-B}}_{min}$ & $C^{\text{B-B-A}}_{min}$ & $C^{\text{A-B-A}}_{min}$ & $C^{\text{A-B-B}}_{min}$ & Attr$_{\text{A-B}}$ & Rep$_{\text{A-B}}$ \\
     \hline
     		Mg-Al & 3.096 & 2.068 & 5.017 & 2.80 & 2.80 & 2.80 & 2.80 & 0.49 & 0.49 & 0.36 & 0.36 & 0.00 & 0.00  \\
     		
     		Mg-Zn & \textbf{3.065} & 1.470 & \textbf{4.960} & 2.80 & 2.00 & 2.80 & 2.80 & \textbf{1.00} & 0.30 & 2.00 & 1.00 & 0.00 & 0.00  \\
     		
     		Al-Zn & 2.856 & 2.840 & 4.57 & 2.80 & 2.80 & 2.80 & 2.80 & 0.36 & 2.00 & 2.00 & 2.00 & 0.075 & 0.075  \\
    \hline \hline
    \end{tabular}
    }%
  \end{center}
\end{table}


\end{document}